\documentclass[preprint,eqsecnum,aps,keywords,showpacs]{revtex4}
\usepackage{dcolumn}
\usepackage{graphicx}
\usepackage{amsmath}
\usepackage{amssymb}
\usepackage{epsfig}
\usepackage{float}
\usepackage{latexsym}
\usepackage{rotating}
\begin{document}

\title{ \small Geodesic Motion in a  Charged 2D Stringy Blackhole Spacetime}
\author{
Rashmi Uniyal\footnote{Electronic address: {\em rashmiuniyal001@gmail.com}}${}^{}$}
\affiliation{ \small{Department of Physics, Gurukul Kangri Vishwavidyalaya,\\ Haridwar 249 407, Uttarakhand, India} }

\author{
Hemwati Nandan\footnote{Electronic address: {\em hnandan@iucaa.ernet.in}}${}^{}$}
\affiliation{Department of Physics, Gurukula Kangri Vishwavidyalaya, Haridwar 249 407, Uttarakhand, India}

\author{
K D Purohit\footnote{Electronic address: {\em kdpurohit@rediffmail.com}}${}^{}$}
\affiliation{Department of Physics, HNB Garhwal University, Srinagar Garhwal 246 174, Uttarakhand, India}

\vspace{2.5cm}
\begin{abstract}
We study the timelike geodesics and geodesic deviation for a two-dimensional
stringy blackhole spacetime in Schwarzschild gauge.
We have analyzed the properties of effective potential along with
the structure of the possible orbits for test particles with different
settings of blackhole parameters.
The exactly solvable geodesic deviation equation is used to obtain corresponding deviation vector.
The nature of deviation and tidal force is also examined in view of the behavior of
corresponding deviation vector.
The results are also compared with an another two-dimensional stringy blackhole spacetime.\\

\end{abstract}


\maketitle
\section{Introduction}	
Lower dimensional theories of gravity provide the simplified contexts to study
various physical phenomena \cite{harvey}.
In particular, the charged dilatonic blackholes (BHs) have been studied extensively
 in the literature \cite{gibbon,gibM,gar,wit,Mcg}
and such BHs have diverse connections to supergravity \cite{gibbon},
 Kaluza-Klein \cite{gibM}, string \cite{gar} and conformal field theories \cite{wit}.\\
The study of geodesics for test particles in given sapcetime
is one way to understand the gravitational field around the black hole
and therefore, to investigate the given spacetime, first step may be to
consider the motion of test particles in it.
Further, the relative acceleration between free test particles
falling in a gravitational field is
given by \textit{geodesic deviation}, which is very useful to understand the physical
effects arising due to gravitational field and thus the geometry of given spacetime \cite{piran}.
The study of geodesics and geodesic deviation is a popular topic of
interest among mathematicians and physicists and
a number of studies related to the geodesic motion in the background of various
black hole spacetimes in general relativity
(GR) \cite{piran,mis,ad,sch,schandra,haw,mabd,gad,pug,lies,grunau,hack,psj,el}
and other alternative theories of gravity \cite{ker,ell,koley,ghosh}
are made time and again.\\
Various aspects concerning the geodesic motion around the spacetimes arising from
string theory (i.e. stringy BHs) in 2D and 4D,
which include energy distribution \cite{ecv},
varying parameters problem \cite{EC}, the
kinematics of time-like geodesic congruences \cite{gp,hn,macdbh} and geodesic
deviation \cite{ell,koley,ghosh,gad} are also widely discussed.\\
The spacetime considered in the present work is of 2D dilatonic blackhole, arising from
heterotic string theory \cite{Mcg}.
We are interested in this particular spacetime because it has many analogies with the
Reissner-Nordstr$\ddot{o}$m black hole in 4D in GR.
The main objective of this paper is to study the geodesic structure for test particles
in above mentioned spacetime and its uncharged counterpart.\\
The structure of the paper is as follows. In the next section, we present a
brief introduction to the model and the line element considered.
In section III, we study the geodesic equations, effective potential and thus the
geodesic structure for timelike geodesics.
We have discussed about the probable orbits by analyzing the properties of effective potential.
Section IV is concerned about the geodesic deviation and
the evolution of deviation vector, for all the possible cases in view of the behavior of effective potential.
In section V, we conclude the results obtained with the future possibilities.

\section{The Model and Spacetime}
Two-dimensional effective action in heterotic string theory \cite{Mcg} is given as,
\begin{equation}
S={\frac{1}{2\pi}}\int{{d^{2}}x{\sqrt{-g}}{e^{-2\phi}}\left[R+4{(\nabla \phi)^{2}}+4{{\lambda}^{2}}-{\frac{1}{2}}{F^{2}}\right]},
\label{eqn1}
\end{equation}
where $\phi$ is the dilaton field,
${{\lambda}^{2}}$ is the cosmological constant and
${F_{\mu \nu}}$ is the Maxwell stress tensor.
The line element for charged two-dimensional blackhole solution \cite{Mcg,ecv} derived
from action given in Eq.(~\ref{eqn1}) (in \textit{Schwarzschild gauge}) is given by,
\begin{equation}
{ds}^2 = -g(r){dt}^{2}+\frac {{dr}^{2}}{g(r) },
\label{eqn2}
\end{equation}
where,
\begin{equation}
g(r) = 1 - \frac{M}{\lambda}e^{-2\lambda r} + \frac{Q^2}{4{\lambda}^2}e^{-4\lambda r},
\label{eqn3}
\end{equation}
 horizons (i.e. event and Cauchy) of above spacetime are given as,
\begin{equation}
r_{\pm} = {\frac{1}{2\lambda}}\ln\left(\frac{Q}{2\lambda(M\pm\sqrt{M^{2}-Q^{2}})}\right).
\label{horizons_ch}
\end{equation}
here, $0 < t < +\infty$, $r_+ < r < +\infty$, $r_+$ is the event horizon of the blackhole.
The parameters $M$ and $Q$ are the mass and electric charge of 2D analogue of
$Reissner$-$Nordstr{\ddot{o}}m$ blackhole spacetime
given by eq.(~\ref{eqn2}).
 The asymptotic flat region is $r=+\infty$ (since Q and $\lambda$ are positive)
 and the curvature singularity is at $r=-\infty$.
 There are two singularities present in the spacetime given by eq.(~\ref{eqn2})
 as mentioned in eq.(~\ref{horizons_ch}).
  For extremal case $M=Q$, both singularities are merged,
  hence there exists only one event horizon.
 It is worth noticing that $M<Q$ represents the case of a naked singularity.\\
 In the absence of charge, i.e. the Maxwell stress tensor ${F_{\mu \nu}}$
 is not present in eq.(~\ref{eqn1}), the resulting action
 now arises as the effective action
 describing the radial modes of extremal dilatonic BHs in
 four or higher dimensions \cite{gibbon,gibM,gar,hor,gid},
where the metric element given in eq.(~\ref{eqn2}) now reads off,
\begin{equation}
g(r) = 1 - \frac{M}{\lambda}e^{-2\lambda r},
\label{eqn6}
\end{equation}
and the position of event horizon is given by,
\begin{equation}
r_{H} = {\frac{1}{2\lambda}}\ln\left(\frac{M}{\lambda}\right).
\label{eqn7}
\end{equation}
where $r_{H} < r < +\infty$.
The timelike geodesic motion in the background geometry of the above mentioned spacetime is discussed below.

\section{Timelike Geodesics and The Effective Potential}
In order to understand the spacetime given in eq.(\ref{eqn2}), let us first consider the behavior of geodesics.
The geodesic equation is given as,
\begin{equation}
{\ddot{x}}^{\mu}+\Gamma^{\mu}_{\nu\rho}{\dot{x}}^{\nu}{\dot{x}}^{\rho}=0
\end{equation}
where $x^{\mu}$ are the coordinates $r$ and $t$, $\mu$ takes values $0$ and $1$ for $t$ and $r$ respectively.
Here $\Gamma^{\mu}_{\nu\rho}$ represents the Christoffel symbols (for non-vanishing components of Christoffel symbols see \textbf{Appendix}).
The geodesic equation for $t$ results as below,
\begin{equation}
\ddot{t} + {\frac{g'(r)}{g(r)}}{\dot r}{\dot t}=0,
\label{eqn7}
\end{equation}
where the prime ($'$) and dot ($\cdot{}$) denote the differentiation with
respect to $r$ and $\tau$ (affine parameter) respectively.
On integrating eq.(\ref{eqn7}) once, we obtain
\begin{equation}
\dot t = \frac{E}{g(r)}.
\label{eqn9}
\end{equation}
where $E$ is a constant of the
motion which arises due to the absence of the coordinate $t$ in the metric coefficients and its value provides a way to understand the orbits of particles in the above mentioned geometry.
The geodesic equation corresponding to $r$ now reads as,
\begin{equation}
 \ddot{r}+\left[{\frac {2\,\lambda\, \left( 2\,M{{\rm e}^{-2\,\lambda\,r}}\lambda-{Q}^
{2}{{\rm e}^{-4\,\lambda\,r}} \right) }{-4\,{\lambda}^{2}+4\,M{{\rm e}
^{-2\,\lambda\,r}}\lambda-{Q}^{2}{{\rm e}^{-4\,\lambda\,r}}}}\right]({\dot{r}}^2 - E^2) = 0.
\label{eqn8}
\end{equation}
Since our main aim is to study the timelike geodesics in the above mentioned spacetime, the motion is constrained as,
\begin{equation}
-g(r){\dot{t}}^2+{g(r)}^{-1}{\dot{r}}^2 = -1\hspace{5mm}\left(g_{\mu\nu}{\dot{x}^{\mu}}{\dot{x}^{\nu}}=-1\right),
\label{eqn10}
\end{equation}
and now using eq.(\ref{eqn9}) in eq.(\ref{eqn10}) for the timelike constraint, we obtain
\begin{equation}
 {\dot{r}}^2 = E^{2}- g(r) = -V(r) + E_{0},
\label{eqn11}
\end{equation}
where $E_{0}$ is a constant and $V(r)$ is the effective potential given by,
\begin{equation}
V(r)=E_{0}-E^{2}+g(r)\equiv E_{0}-E^{2}+1-{\frac{M}{\lambda}}e^{-2\lambda r}+{\frac{Q^{2}}{4{\lambda}^{2}}}e^{-4\lambda r}.
\label{p_eff}
\end{equation}
The condition for circular orbit (i.e. ${dV(r)}/{dr}=0$), which locates the maxima and minima of the effective potential energy curve \cite{hartle,Poisson,wald}, out of which minima corresponds to the position of stable circular orbit.
Differentiating eq.({\ref{p_eff}}) with respect to $r$ once and equating it to zero, the radius of stable circular orbit is given as,
\begin{equation}
{r_c}={\frac{1}{2\lambda}}\ln\left(\frac{Q^{2}}{2M\lambda}\right).
\label{r_c}
\end{equation}
Similar to classical electron radius, classical charge radius is defined as
${r_{classical}}={Q^2}/M$\cite{pug,bin}. From eq.({\ref{r_c}}) relation between $r_c$ and $r_{classical}$ can be obtained as,
\begin{equation}
{r_c}={\frac{1}{2\lambda}}\ln\left(\frac{r_{classical}}{2\lambda}\right).
\label{eq:r_cl}
\end{equation}
Such condition for allowed timelike circular geodesics in case of $4D$ $Reissner$-$Nordstr{\ddot{o}}m$ blackhole spacetime is given by, ${r_c}\geq{r_{classical}}$\cite{pug}. \\
The corresponding condition for spacetime given in eq.({\ref{eqn3}}) comes out to be,
\begin{equation}
\frac{r_c}{\exp(2\lambda {r_c})}\geq 2\lambda.
\label{eq:con_r_c}
\end{equation}
Equivalently, in terms of classical charge radius, the above condition can be expressed as follows,
\begin{equation}
r_{classical}\geq2\lambda\exp\left(2\lambda{r_{classical}}\right).
\label{eq:con_r_general}
\end{equation}
Hence for given Mass (M) and Charge (Q) of the blackhole, relations given in eq. (\ref{eq:con_r_c}) and eq.(\ref{eq:con_r_general}) restricts the value of cosmological constant, $\lambda$ for each case in view to obtain allowed circular timelike geodesics.
As $Q/M$ ratio increases, bound on $\lambda$ shifts towards less positive value.\\
The last stable circular orbit (lsco) i.e. the minimum radius for stable circular orbits is determined by the inflection points of the effective potential $V(r)$, which are obtained by the condition ${d^2}V(r)/d{r^2}$=0.
Applying this condition on effective potential $V(r)$,
\begin{equation}
r_{lsco}={\frac{1}{2\lambda}}\ln\left({\frac{Q^2}{M}}\right),
\end{equation}
or in terms of $r_{classical}$
\begin{equation}
r_{lsco}={\frac{1}{2\lambda}}\ln\left(r_{classical}\right).
\end{equation}
With given values of Mass ($M$), Charge ($Q$) of black hole and the value of cosmological constant ($\lambda$) as obtained from eq.(\ref{eq:con_r_general}), it is observed that $r_{lsco}$ increases as charge dominates over mass. From the physical view point, it supports the nature of effective potential as discussed in the next section, as it becomes repulsive as ratio $Q/M$ increases. \\
Hence, the radius of last stable circular orbit also shifts towards a larger value as charge dominates due to the repulsive nature of effective potential.
The effective potential for geodesic motion corresponding to different regions of a constant of motion is discussed in the next section.

\subsection{Nature of Effective Potential and Structure of Orbits}
In order to visualize the behavior of the
 effective potential given by eq.(\ref{p_eff}) with
different settings of blackhole parameters (i.e. cosmological constant $\lambda$, mass $M$ and charge
$Q$), the representative plots for four different cases are given below.\\
\begin{figure}[h]
\centerline{\includegraphics[scale=0.4]{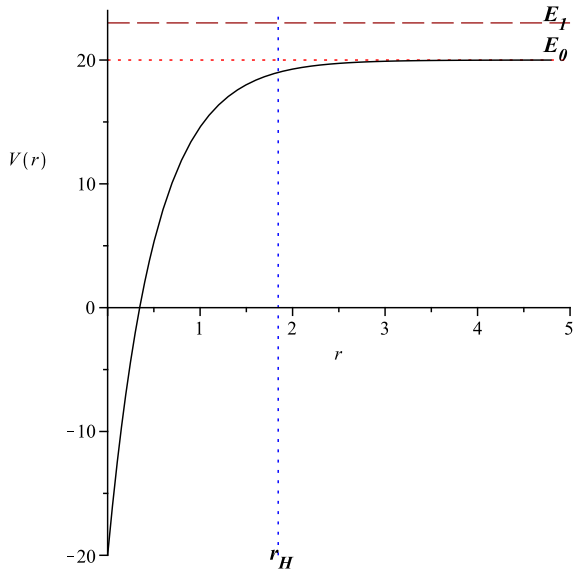}}
\vspace*{8pt}
\caption{Effective potential for $Q=0$ $(M=40,E=1,{E_0}=20,\lambda=1)$, vertical dashed line shows the position of event horizon ${r_H}$ and horizontal dotted line shows the position of $E_0$. \protect\label{f1}}
\end{figure}

\begin{figure}[h]
\centerline{\includegraphics[scale=0.37]{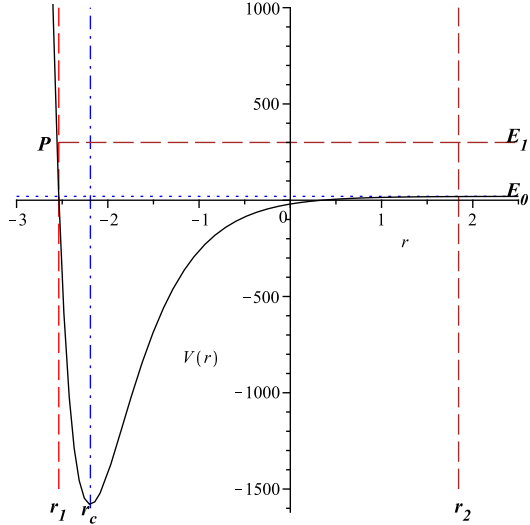}}
\vspace*{8pt}
\caption{Effective potential for $M>Q$ $(M=40,Q=1,E=1,{E_0}=20,\lambda=1)$, dashed vertical lines show the positions of event horizon $r_1$ and cauchy horizon $r_2$, dotdashed vertical line shows the radius of stable circular orbit $r_c$ and horizontal dotted line shows the position of $E_0$. \protect\label{f2}}
\end{figure}

\begin{figure}[h]
\centerline{\includegraphics[scale=0.37]{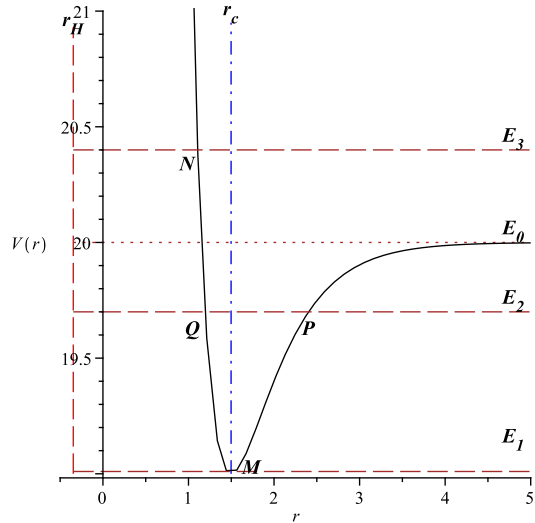}}
\vspace*{8pt}
\caption{Effective potential for $M=Q$ $(M=40,Q=40,E=1,{E_0}=20,\lambda=1)$, dashed vertical line shows the position of event horizon $r_H$,
dotdashed vertical line shows the radius of stable circular orbit $r_c$ for test particle. \protect\label{f3}}
\end{figure}

\begin{figure}[h]
\centerline{\includegraphics[scale=0.4]{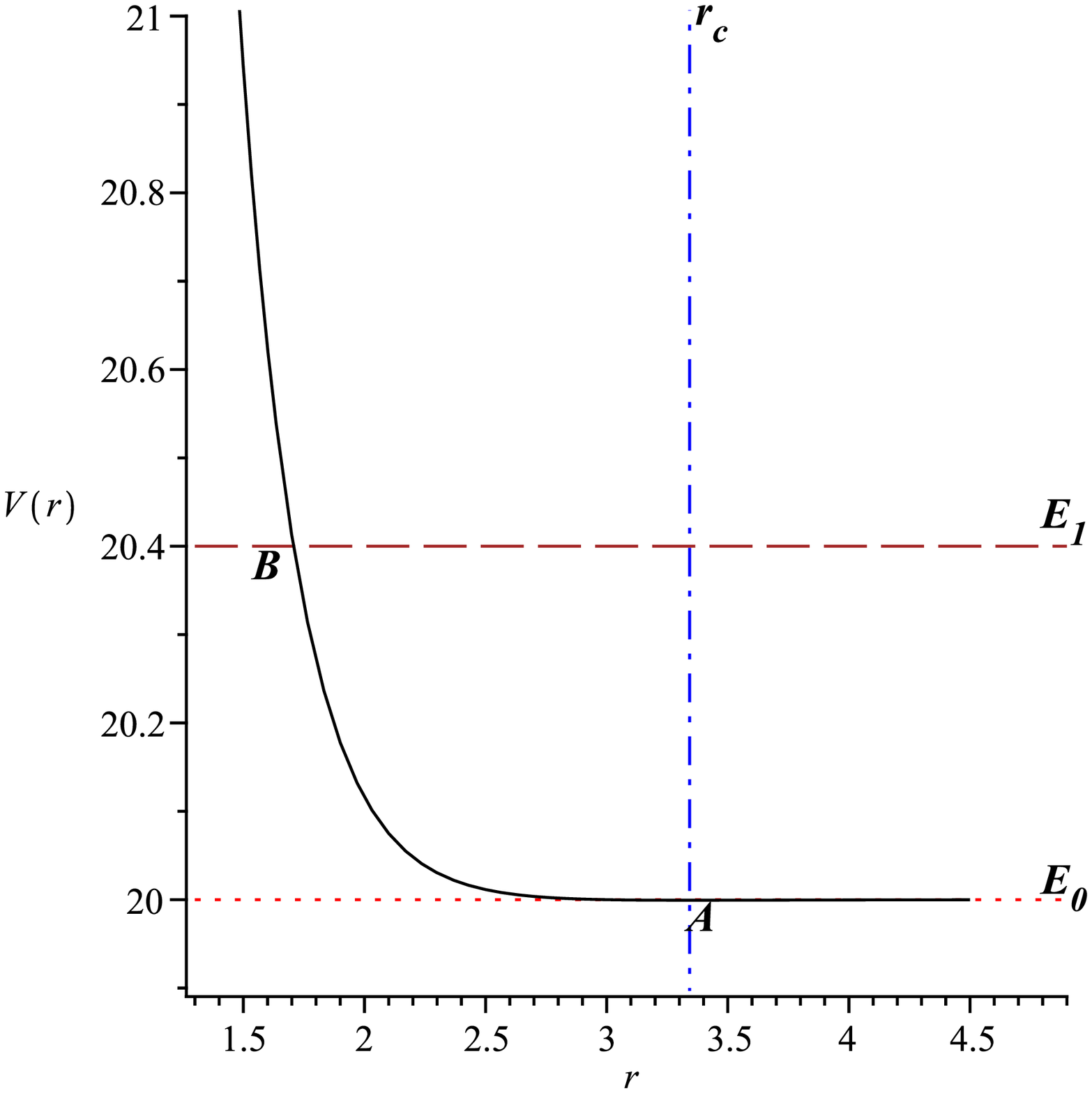}}
\vspace*{8pt}
\caption{Effective potential for naked singularity, where $M<Q$ $(M=1,Q=40,E=1,{E_0}=20,\lambda=1)$, dotdashed vertical line shows the radius of stable circular orbit $r_c$
for test particle. \protect\label{f4}}
\end{figure}

\textbf{Case I}\hskip2mm
In case of uncharged black hole the effective potential as shown in figure (\ref{f1}) is such that $V(r)\leq{E_0}$ always.
If incoming test particle has energy $E_1$ such that ${E_1}>{E_0}$ (see figure (\ref{f1})), the
particle starts from infinity and always leads to the physical singularity situated at $-\infty$, on crossing the horizon.
Hence for an incoming test particle with energy $E>{E_0}$, only terminating orbits are possible in this case.
Even if the particle has energy $E<{E_0}$, it does not seem to have bound orbits in given two dimensional spacetime.\\

While in case of its charged counterpart, nature of potential becomes repulsive as we proceed towards singularity due to presence of charge.
We can compare from figures (\ref{f1}-\ref{f4}) that as charge dominates over mass, this repulsive nature becomes more strong.
One important difference from uncharged case is that, here due to the repulsive nature of potential one can not expect the terminating orbits for test particle.
Hence, the potential barrier due to charge prevents test particle from being dropped into the singularity.
The orbits possible here may be bound (if $E<{E_0}$) and escaping (if $E>{E_0}$).\\

\textbf{Case II}\hskip2mm
For $M>Q$, position of horizons is given in eq.({\ref{horizons_ch}}),
while $r_c$ is given in eq(\ref{r_c}) and $r_{-}<r_{c}<r_{+}$ shows the stable bound circular motion of test particle between the horizons,
as it locates at the minima of potential (see figure(\ref{f2})).
This region between horizons is forbidden in available literature, the particle rather would return quite before reaching $r_c$ as turning point exists where $V(r)=0$ (in this case turning point is just adjacent to the cauchy horizon).
This behavior of the effective potential can be better checked by analyzing corresponding $4D$ spacetime metric in future.
On the other hand, particle with energy ${E_1}>{E_0}$ (as shown in figure(\ref{f2})) starting from infinity would strike at point $P$ and then will return to infinity.
Hence incoming test particle with energy $E>{E_0}$ will have escaping orbit.\\

\textbf{Case III}\hskip2mm
In extremal case when $M=Q$, both horizons coincide at ${r_H}={\frac{1}{2\lambda}}\ln({\frac{Q}{2M\lambda}})$  (see figure(\ref{f3})).
As shown in the figure(\ref{f3}), depending on initial energy of test particle, it may have following kinds of orbits:\\
If test particle has energy $E={E_1}$, it will have a stable circular orbit at point $M$ with radius $r_c$ given in eq.({\ref{r_c}}).
If test particle has energy $E={E_2}$, it will have a bound orbit in which the particle moves between two turning points marked as $P$ and $Q$ in the figure.
If test particle has energy $E={E_3}$, starting from infinity towards central singularity, it will turn back to infinity at point $N$. Hence, it will have an escape orbit in this case.
In any of the above cases, particle coming from infinity will not be able to reach the horizon as $r_H$ lies in the forbidden region.
Hence the possible orbits here are either bound (between two turning points), stable circular or escape.
There are no terminating orbits possible here.\\

\textbf{Case IV}\hskip2mm
For naked singularity with $M<Q$, no horizons exist (see figure(\ref{f4})). If we compare figure(\ref{f3}) and figure(\ref{f4}), we observe that
the position of $r_c$ shifts towards right.
The possible orbits here may therefore be either circular (if test particle energy $E={E_0}$) or escape (if test particle energy $E={E_1}$).\\

\subsection{Analytic Solutions of Geodesic Equations}
Geodesic equation (\ref{eqn9}) and timelike constraint given in eq.(\ref{eqn10}) are used to obtain the solution for $r(\tau)$ and subsequently $t(\tau)$ (for $E=1$) as below,
\begin{equation}
r(\tau) = {\frac{1}{2\lambda}}\ln\left[M \lambda {{\tau}^{2}}+{\frac{Q^{2}}{4M\lambda}}\right],
\label{eqn14}
\end{equation}
\begin{figure}[ht]
  \centerline{\includegraphics[scale=0.2]{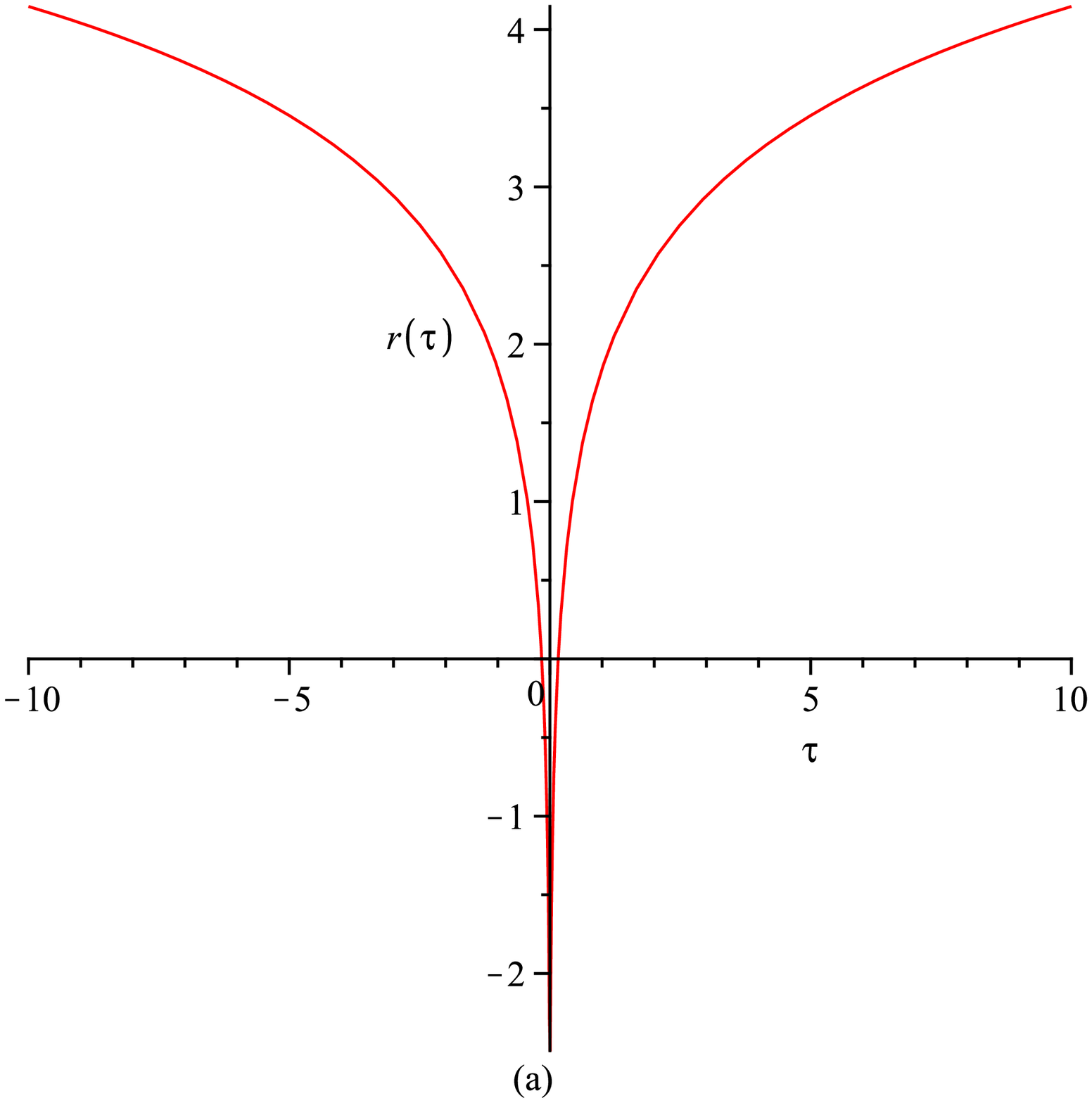}
    \hskip1mm \includegraphics[scale=0.2]{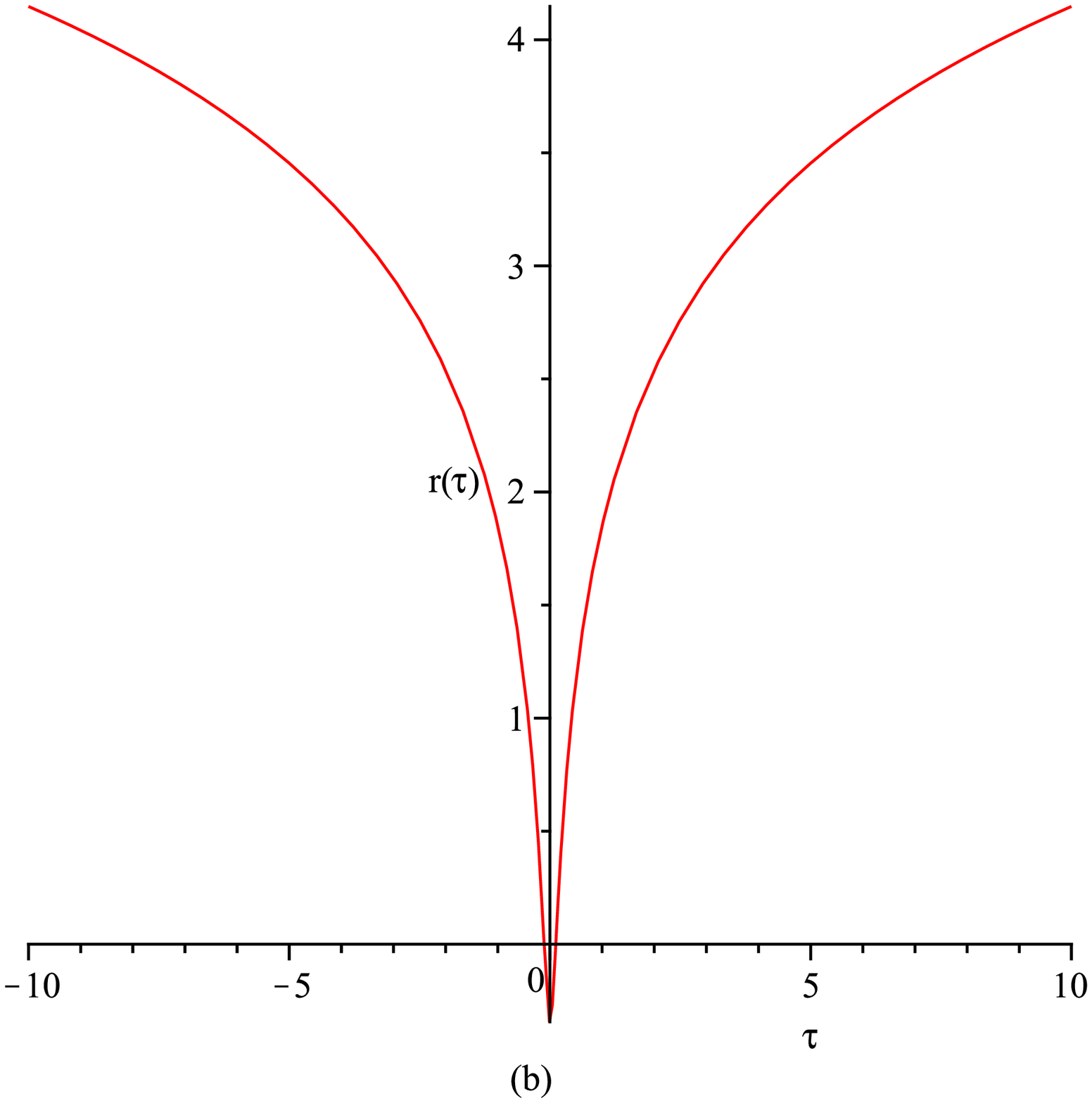}
    \hskip1mm \includegraphics[scale=0.2]{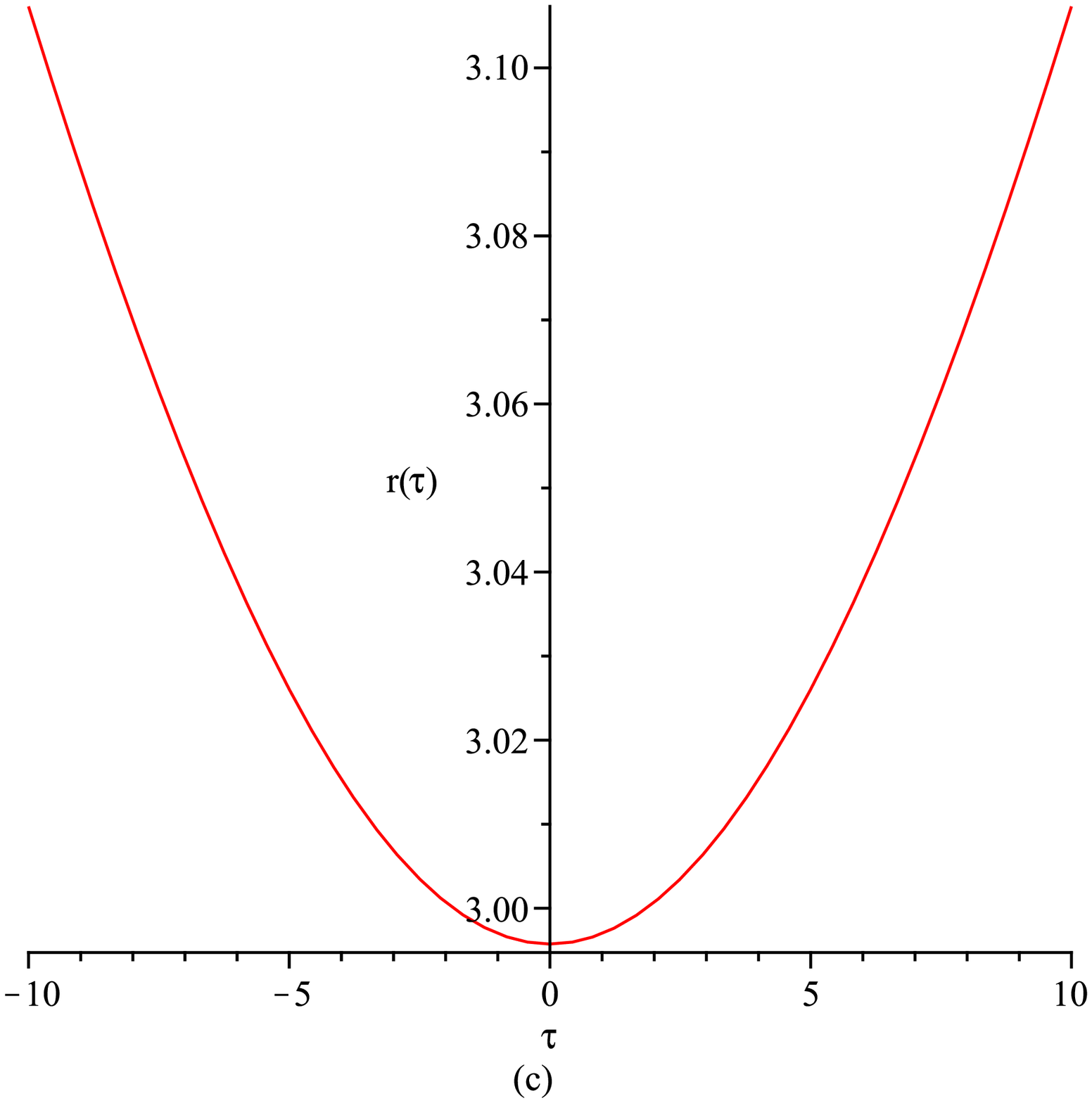}}
\vspace*{8pt}
\caption{
  {Variation of $r(\tau)$,\hspace{2mm}(a) for $M>Q$ ($M=40$,$Q=1$,$\lambda=1$),
  (b) for extremal case $M=Q$  ($M=Q=40$,$\lambda=1$),
  (c) for $M<Q$ ($M=1$,$Q=40$,$\lambda=1$). }}
\protect\label{f5}
\end{figure}
\begin{equation}
t\left(\tau\right) = E\tau + {\frac{E}{4\lambda}}\ln{\left(\frac{{Q^2}-4{M^2}\lambda\tau+4{M^2}{\lambda^2}{\tau^2}}{{Q^2}+4{M^2}\lambda\tau+4{M^2}{\lambda^2}{\tau^2}}\right)}
+\,A \left[{tan}^{-1}\left(\frac{M+2M\lambda\tau}{\sqrt{{Q^2}-{M^2}}}\right)\right]\nonumber
\end{equation}
\begin{equation}
+A \left[\,{tan}^{-1}\left(\frac{-M+2M\lambda\tau}{\sqrt{{Q^2}-{M^2}}}\right)\right],
\label{eqn15}
\end{equation}
where, $A={\frac{E{M^2}}{2M\lambda{\sqrt{{Q^2}-{M^2}}}}}$.
For $Q=0$ these relations reduce to the following,
\begin{equation}
r(\tau) = {\frac{1}{2\lambda}}\ln\left({M\lambda}{{\tau}^{2}}\right),
\label{eqn16}
\end{equation}
\begin{equation}
t \left( \tau \right) =E \tau +{\frac{E}{2\lambda}}\ln\left[{\frac{\lambda\tau-1}{\lambda\tau+1}}\right] .
\label{eqn17}
\end{equation}
\begin{figure}[h]
\centerline{\includegraphics[scale=0.35]{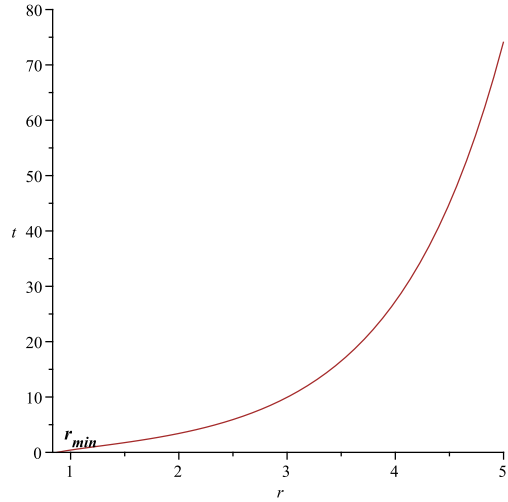}}
\vspace*{8pt}
\caption{Variation of $r$ with $t$ for uncharged case ($E=1$,$M=40$,$\lambda=1$), which shows that $r$ increases exponentially
 with $t$ and ${r_{min}}>{r_H}$}
\protect\label{f8}
\end{figure}
The above figure(\ref{f8}) shows the actual trajectory of particle as a function of time.
It shows that trajectory of uncharged blackhole can extend from $r_H$ to $\infty$, where $r_H$ is the event horizon of uncharged blackhole.

\section{Geodesic Deviation Equation and Tidal Effects}
Given a geodesic curve, we identify the tangent and normal to it as the vectors $e^{\mu}$ and $n^{\mu}$ respectively which satisfy the orthonormality conditions as below,
\begin{equation}
{g_{\mu\nu}}{e^{\mu}}{e^{\nu}}=-1,
\hspace{1cm} {g_{\mu\nu}}{n^{\mu}}{n^{\nu}}=1,
\hspace{1cm} {g_{\mu\nu}}{e^{\mu}}{n^{\nu}}=0.
\label{onormalc}
\end{equation}
Let us assume\cite{koley},
\begin{equation}
{e^{\mu}\equiv\left(\hspace{1mm}\dot{t},\hspace{1mm}\dot{r}\hspace{1mm}\right)},
\hspace{1cm}
{n^{\mu}\equiv\left(\hspace{1mm}f(r)\dot{r},\hspace{1mm} h(r)\dot{t}\hspace{1mm}\right)}
\end{equation}
and substituting it in orthonormality condition given in eq.(\ref{onormalc}), we obtain:
\begin{equation}
{f(r)=\sqrt{\frac{g_{11}}{-g_{00}}}},
\hspace{1cm} {h(r)=\sqrt{\frac{-g_{00}}{g_{11}}}}
\end{equation}
%$f(r)=\sqrt{\frac{g_{11}}{-g_{00}}}$ \hspace{1cm} $h(r)=\sqrt{\frac{-g_{00}}{g_{11}}}$\\
which leads to the temporal and spatial components of the normal vector as follows:
\begin{equation}
n^{0}=\frac{\dot{r}}{g(r)}, \hspace{2cm} n^{1}=\dot{t}g(r).
\label{eqn19}
\end{equation}
where $g(r)$ is given by eq.(\ref{eqn3}).
Equation for normal deformations (where $\eta^{\mu}={\eta}n^{\mu}$,$\eta^{\mu}$ is deviation vector) {\cite{hartle,Poisson,wald}}:\\
\begin{equation}
\ddot{\eta} \left( \tau \right) + {R_{\mu \nu \rho \sigma}}{e^{\mu}}{n^{\nu}}{e^{\rho}}{n^{\sigma}} \eta \left( \tau \right)=0,
\label{eqn20}
\end{equation}

If we substitute the Riemann tensor components (see \textbf{Appendix}) for the two-dimensional stringy blackhole line
element and use the normal and tangent mentioned above, the deviation
equation is given as \\
\begin{equation}
\ddot{\eta} \left( \tau \right)+\frac{8{M^2}{\lambda^2}(4{M^2}{\lambda^2}{\tau^2}-3Q^2)}{(4{M^2}{\lambda^2}{\tau^2}+Q^2)^2}\eta(\tau)=0,
\label{gdev}
\end{equation}
the above deviation equation gives the solution in form of complex Legendary functions. As deformation vector should be real, we can take the range of affine parameter for which real solutions exist. Hence in eq.({\ref{gdev}}), if the coefficient of $\eta(\tau)$ in second term becomes negative, solutions for $\eta$ are real within a specified range of affine parameter $\tau$ such that, ${{\tau}^2}<\frac{3{Q^2}}{4{M^2}{{\lambda}^2}}$. With this condition, deviation equation given in eq.({\ref{gdev}}) leads to the following solution,
\begin{equation}
\eta \left( \tau \right) ={\frac {{\it C_1}\,\tau}{4\,{M}^{2}{\lambda
}^{2}{\tau}^{2}+{Q}^{2}}}+{\frac {{\it C_2}\, \left( -3\,{Q}^{4}+24\,
{M}^{2}{\lambda}^{2}{\tau}^{2}{Q}^{2}+16\,{M}^{4}{\lambda}^{4}{\tau}^{
4} \right) }{4\,{M}^{2}{\lambda}^{2}{\tau}^{2}+{Q}^{2}}},
\label{eta}
\end{equation}
where the integration constants ${C_1}$ and ${C_2}$ are given as follows,
\begin{equation}
{C_1}=\frac{{{\eta}_0}}{{{\tau}_0}}\left(4{{{\tau}_0}^2}{M^2}{{\lambda}^2}+{Q^2}\right)-\frac{C_2}{{\tau}_0}\left(16{{{\tau}_0}^4}{M^4}{{{\lambda}^4}}+24{{{\tau}_0}^2}{M^2}{{\lambda}^2}{Q^2}-3{Q^4}\right),
\end{equation}
\begin{equation}
{C_2}=\frac{{{\dot{\eta}}_0}{{\tau}_0}\left(4{{{\tau}_0}^2}{M^2}{{\lambda}^2}+{Q^2}\right)+4{{\eta}_0}{{{\tau}_0}^2}{M^2}{{{\lambda}^2}}-{{\eta}_0}{Q^2}}{48{{{\tau}_0}^4}{M^4}{{{\lambda}^4}}+24{{{\tau}_0}^2}{M^2}{{\lambda}^2}{Q^2}+3{Q^4}}.
\end{equation}
where ${{\eta}_0}$
and ${{{\dot\eta}}_0}$ are normal deformation and its first derivative respectively, at initial point ${\tau}_0$.
If we choose the initial condition as ${{\tau}_0}=0$, the integration constants transform as ${C_1}={Q^2}{{{\dot\eta}_0}}$ and ${C_2}={-{\eta}_0}/{3{Q^2}}$.
For $M>>Q$, deviation equation takes the following form in limit $Q\rightarrow0$ (which thus represents the case of uncharged black hole as well) (with condition ${\tau}<0$),
\begin{equation}
\ddot{\eta}\left( \tau \right) -2\,{\frac {\eta
 \left( \tau \right) }{{\tau}^{2}}}=0,
\label{gdev_u}
\end{equation}
Solution for $\eta$ given as,
\begin{equation}
\eta \left( \tau \right) ={D_1}\,{\tau}^{2}+{\frac {{D_2}}{
\tau}},
\label{eta_M}
\end{equation}
where the integration constants $D_1$, $D_2$ are given as follows,
\begin{eqnarray}
{D_1}=-{\frac{{\tau}_0}{3}}({{{\dot\eta}_0}}{{\tau}_0}-2{{\eta}_0}),\\
{D_2}={\frac{1}{3{{\tau}_0}^2}}({{{\dot\eta}_0}}{{\tau}_0}+{{\eta}_0}).
\end{eqnarray}
where ${{\eta}_0}$
and ${{{\dot\eta}}_0}$ are normal deformation and its first derivative respectively,
at initial point ${\tau}_0$.\\

After obtaining $\eta$ for each case, deviation vector is given by ${\eta}^{\mu}={\eta}{n^{\mu}}$ (where $\mu=0,1$).
For $Q=0$ and $M>>Q$, following the condition for affine parameter $\tau$ as ${\tau}<0$, the range of $\tau$ (for $r>r_H$) is $-\infty<\tau<0$.
If we consider a pair of neighboring geodesics at two adjacent values of $\tau$ and assume $\dot{\eta}$ to
be positive, then from eq.(\ref{eta_M}) we have to choose the solution for $\eta$ that is propotional
to ${\tau}^2$. The evolution of $\eta$ then suggests that as we move further to a smaller value of $\tau$
in above range, the geodesics spread out (diverge).
On the other hand, if we assume $\dot{\eta}$ initially negative we have to choose the solution for $\eta$ that is propotional to $\tau^{-1}$, the geodesics converge atleast locally. It means that if we see just the neighboring geodesics $\eta$ decreases but becomes constant after sometime. As we go to smaller $\tau$, $\dot{\eta}$ becomes smaller and approaches to zero as $\tau\rightarrow{-\infty}$ ($r\rightarrow{\infty}$). Therefore, finally the geodesics become parallel to each other.\\
A similar analysis can be carried out for other cases with $M=Q$ and $M<<Q$.
From the condition on affine parameter $\tau$ and solution for $\eta(\tau)$ given in eq.(\ref{eta}), relative motion of neighboring geodesics can be visualized.
With an initial value of $\eta$, if $\dot{\eta}$ is positive we will have to choose the solution that is propotional to the increasing power of $\tau$ whereas for negative $\dot{\eta}$ we will have to choose the solution that is propotional to the decreasing power of $\tau$.
It means that if geodesics are initially flying apart they will diverge further, atleast in their nearest neighborhood.
While if geodesics are initially coming towards each other, they converge if we see them relative to neighboring geodesics.
The difference from previous case here is that with initially negative deformation, $\eta$ keeps on decreasing with $\tau$, it does not become constant at smaller values of $\tau$ as before.\\
The tidal gravitational force acts to change the separation between a pair of neighboring geodesics and
we can substitute the value of $\tau$ from the expression of $r(\tau)$ in the geodesic deviation eq.(\ref{gdev}) and eq.(\ref{gdev_u}), to analyze the nature of the tidal force. Let us consider the case of $Q\rightarrow0$, (following the solutions obtained for $\eta$ in eq.(\ref{eta_M})) for $\eta={\tau}^2$, the relative force is constant throughout (i.e. $\ddot{\eta}$ is a constant when $\eta={\tau}^2$). For $\eta={\tau}^{-1}$, the force vanishes as $r\to\infty$ which implies that the geometry is flat in this limit and the geodesics will become parallel to each other, whereas for $r\rightarrow{r_H}$, the tidal force has a constant value.
Similarly, for the generalized case, the solutions obtained for $\eta$ in eq.(\ref{eta}) are used for the analysis of the tidal effects.
For the solution of $\eta$ that is propotional to ${\tau}^2$, the corresponding tidal gravitational force becomes constant as $\tau\rightarrow\infty$.
While for $\eta$ propotional to ${\tau}^{-1}$, tidal force vanishes as $\tau\rightarrow\infty$ ($r\rightarrow\infty$) and geodesics become parallel as mentioned above.\\
Another important point to mention here is that in two dimensions, timelike geodesics converge to a point (i.e. focus) within a finite value of affine parameter if, $R\leq0$ \cite{skgeo} (where $R$ is the Ricci Scalar in given geometry of spacetime).
For the cases where, $Q=0$ and $M>>Q$, Ricci scalar is non-negative (it tends to zero when $r\rightarrow\infty$).
Hence the geodesics do not seem to converge to a point within a finite range of affine parameter $\tau$.
However for the cases $M=Q$ and $M<<Q$ Ricci scalar is negative, it tends to zero with $r\rightarrow\infty$ such that the geodesics may converge to a point within a finite range of affine parameter $\tau$.

\section{Discussion and Conclusions}
In the present work, we have studied geodesics of  2D dilatonic charged and uncharged black holes.
In view of the nature of the effective potential,
it is observed that in case of charged blackhole there exists a potential barrier due to the presence of charge, which does not allow the singularity to swallow the test particle.
Bound as well as unbound orbits are found to exist for a test particle in case of charged blackhole,
while only terminating orbits are possible for its uncharged counterpart.\\

%We have also confirmed the fact that the main factors effecting the nature of effective potential are mass $M$ and %charge $Q$ of the black hole, if cosmological constant $\lambda$ is fixed.\\
In view of the evolution of the deformation vector alogwith energy condition (i.e. condition on Ricci Scalar), we conclude that for the cases $M=Q$ and $M<<Q$ (which represents a naked singularity case), the geodesic focusing is possible.
However in case of $M>>Q$ and $Q=0$, geodesics may either diverge or converge depending on initial condition (whether $\dot{\eta}(\tau)$ is positive or negative initially) and finally become parallel as $\dot{\eta}(\tau)$ approaches zero.
If there is geodesic focusing in any case, it will disappear completely once mass dominates over charge.
The results obtained when compared with \cite{koley} also confirms the absence of geodesic focusing for the cases when $M>>Q$ and $Q=0$.
The geodesic focusing and defocusing for the above mentioned cases can be better visualized by investigating the solution of Raychaudhuri equation for expansion scalar.\\
For more accurate visualization of the geodesic motion in the background geometry of stringy blackhole spacetime we intend to report on this in future by using 4D extension of such black holes in heterotic string theory.

\section*{Acknowledgments}
The authors are indebted to  the anonymous referee for useful suggestions and comments  on the manuscript which helped us to improve the presentation of the paper significantly.
One of the authors HN would like to thank Department of Science and Technology, New Delhi for financial support through grant no. SR/FTP/PS-31/2009 and IUCAA, Pune for support under its visiting associateship program.
RU acknowledges the  support from CTP,  JMI, New Delhi under its visitors program.
The authors are also thankful to Buddhi V Tripathi for useful discussions.

\section*{Appendix}

Non-zero components of the Christoffel symbols of second kind for the line element given in eq.(\ref{eqn2}):
\begin{eqnarray}	%A.1
\begin{array}{rcl}
{{\Gamma}^{0}_{01}}=-2\,{\frac {\lambda\, \left( 2\,M{{\rm e}^{-2\,\lambda\,r}}\lambda-{Q}
^{2}{{\rm e}^{-4\,\lambda\,r}} \right) }{-4\,{\lambda}^{2}+4\,M{
{\rm e}^{-2\,\lambda\,r}}\lambda-{Q}^{2}{{\rm e}^{-4\,\lambda\,r}}}},\\[8pt]
{{\Gamma}^{1}_{00}}=  -1/8\,{\frac { \left( -4\,{\lambda}^{2}+4\,M{{\rm e}^{-2\,\lambda\,r}}
\lambda-{Q}^{2}{{\rm e}^{-4\,\lambda\,r}} \right)  \left( 2\,M{{\rm e}
^{-2\,\lambda\,r}}\lambda-{Q}^{2}{{\rm e}^{-4\,\lambda\,r}} \right) }{
{\lambda}^{3}}},\\[8pt]
{{\Gamma}^{1}_{11}}=  2\,{\frac {\lambda\, \left( 2\,M{{\rm e}^{-2\,\lambda\,r}}\lambda-{Q}^
{2}{{\rm e}^{-4\,\lambda\,r}} \right) }{-4\,{\lambda}^{2}+4\,M{{\rm e}
^{-2\,\lambda\,r}}\lambda-{Q}^{2}{{\rm e}^{-4\,\lambda\,r}}}}.
\end{array}
\label{app1}
\nonumber
\end{eqnarray}
Non-zero Ricci tensor components in the coordinate frame $(R_{\mu \nu}={R^{\alpha}}_{\mu \alpha \nu})$ are :
\begin{eqnarray}	%A.2
\begin{array}{rcl}
R_{00} = -{\frac { \left( M{{\rm e}^{-2\,\lambda\,r}}\lambda-{Q}^{2}{
{\rm e}^{-4\,\lambda\,r}} \right)  \left( -4\,{\lambda}^{2}+4\,M{
{\rm e}^{-2\,\lambda\,r}}\lambda-{Q}^{2}{{\rm e}^{-4\,\lambda\,r}}
 \right) }{2{\lambda}^{2}}},\\[8pt]
R_{11} = {\frac {8{\lambda}^{2} \left( M{{\rm e}^{-2\,\lambda\,r}}\lambda-{Q}
^{2}{{\rm e}^{-4\,\lambda\,r}} \right) }{-4\,{\lambda}^{2}+4\,M{
{\rm e}^{-2\,\lambda\,r}}\lambda-{Q}^{2}{{\rm e}^{-4\,\lambda\,r}}}}.
\end{array}
\label{app2}
\nonumber
\end{eqnarray}
Non-zero components of Riemann tensor :
\begin{eqnarray}	%A.3
\begin{array}{rcl}
R_{0101} = R_{1010} = 2\,M\lambda{{\rm e}^{-2\,\lambda\,r}}-2\,{Q}^{2}{{\rm e}^{-4\,\lambda
\,r}},\\[8pt]
R_{0110} = R_{1001} = -2\,M\lambda{{\rm e}^{-2\,\lambda\,r}}+2\,{Q}^{2}{{\rm e}^{-4\,\lambda
\,r}}.
\end{array}
\label{app3}
\nonumber
\end{eqnarray}
Ricci scalar :
\begin{equation}
R=4\,M{{\rm e}^{-2\,\lambda\,r}}\lambda-4\,{Q}^{2}{{\rm e}^{-4\,\lambda
\,r}}.
\nonumber
\end{equation}

\end{document}